\documentclass[pdflatex,sn-basic]{sn-jnl}

\usepackage{graphicx}%
\usepackage{multirow}%
\usepackage{amsmath,amssymb,amsfonts}%
\usepackage{amsthm}%
\usepackage{mathrsfs}%
\usepackage[title]{appendix}%
\usepackage{xcolor}%
\usepackage{textcomp}%
\usepackage{manyfoot}%
\usepackage{booktabs}%
\usepackage{algorithm}%
\usepackage{algorithmicx}%
\usepackage{algpseudocode}%
\usepackage{listings}%
\usepackage{caption}%
\usepackage{CJKutf8}
\usepackage{pdfpages}
\usepackage{subcaption}

\usepackage{geometry}

\begin{document}
\begin{CJK}{UTF8}{gbsn}

\title[Article Title]{DocReLM: Mastering Document Retrieval with Language Model}


\author[1,2]{\fnm{Gengchen} \sur{Wei}}\email{weigengchen@pjlab.org.cn}
\equalcont{These authors contributed equally to this work.}

\author[1,3]{\fnm{Xinle} \sur{Pang}}\email{pangxinle@pjlab.org.cn}
\equalcont{These authors contributed equally to this work.}

\author[1,4]{\fnm{Tianning} \sur{Zhang}}\email{zhangtianning@pjlab.org.cn}
\equalcont{These authors contributed equally to this work.}

\author[1,5]{\fnm{Yu} \sur{Sun}}\email{sunyu@pjlab.org.cn}

\author*[1]{\fnm{Xun} \sur{Qian}}\email{qianxun@pjlab.org.cn}
\author*[1,6]{\fnm{Chen} \sur{Lin}}\email{linchen@pjlab.org.cn}
\author*[1]{\fnm{Han-Sen} \sur{Zhong}}\email{zhonghansen@pjlab.org.cn}
\author[1]{\fnm{Wanli} \sur{Ouyang}}\email{ouyangwanli@pjlab.org.cn}

\affil*[1]{\orgname{Shanghai Artificial Intelligence Laboratory}, \orgaddress{\city{Shanghai}, \postcode{200232}, \country{China}}}
\affil[2]{\orgname{
Fudan University}, \orgaddress{\city{Shanghai}, \postcode{200433}, \country{China}}}
\affil[3]{\orgname{
Harbin Institute of Technology}, \orgaddress{\city{Harbin}, \postcode{150006}, \country{China}}}
\affil[4]{\orgname{The Chinese University of Hong Kong}, \orgaddress{\city{Hong Kong}, \postcode{999077}, \country{China}}}
\affil[5]{\orgname{Shanghai Jiaotong University}, \orgaddress{\city{Shanghai}, \postcode{200030}, \country{China}}}
\affil[6]{\orgname{University of Oxford}, \orgaddress{\city{Oxford}}, \postcode{OX1 2JD}, \country{United Kingdom}}



\abstract{

With over 200 million published academic documents and millions of new documents being written each year, academic researchers face the challenge of searching for information within this vast corpus. However, existing retrieval systems struggle to understand the semantics and domain knowledge present in academic papers. In this work, we demonstrate that by utilizing large language models, a document retrieval system can achieve advanced semantic understanding capabilities, significantly outperforming existing systems.
 Our approach involves training the retriever and reranker using domain-specific data generated by large language models. Additionally, we utilize large language models to identify candidates from the references of retrieved papers to further enhance the performance. We use a test set annotated by academic researchers in the fields of quantum physics and computer vision to evaluate our system's performance. The results show that DocReLM achieves a Top 10 accuracy of 44.12\% in computer vision, compared to Google Scholar's 15.69\%, and an increase to 36.21\% in quantum physics, while that of Google Scholar is 12.96\%. 

}




\maketitle

\section{Introduction}\label{sec_introduction}



According to Crossref, the academic field contains over 140 million pieces of academic literature, with millions of new papers published each year (\cite{2023-crossref}).
 This exponential growth has made it increasingly challenging for researchers to stay updated with the latest developments in their fields. Consequently, there is a growing demand for efficient semantic-based document retrieval systems(\cite{10.1145/3486250}). However, developing such a system poses significant challenges(\cite{vladika-matthes-2023-scientific}). Firstly, academic papers are written in a highly specialized language with domain-specific knowledge, necessitating a deep understanding of the context to extract relevant information(\cite{wadden-etal-2020-fact}). Secondly, comprehending a paper involves not only reading its content but also understanding the references and their relationships. This requires the system to be aware of the interconnections among references within the paper. Lastly, users may not have precise keywords to describe their information needs, requiring the system to understand the user's intent and provide the most relevant papers accordingly. Though some tries (\cite{searchthearxiv2024}) have been made to create a document retrieval system with a general embedding model, there is still a gap between the models' training task and the real searching. Therefore the performance is not good.

In this study, we introduce DocReLM, a \textbf{Doc}ument \textbf{Re}trieval system enhanced with \textbf{L}anguage \textbf{M}odels to solve these challenges. LLMs (\cite{devlin-etal-2019-bert, gpt-1}, currently among the most prominent areas of focus in natural language processing, demonstrate robust capabilities in understanding complex linguistic constructs. Their application across various domains has shown continuous improvement(\cite{zhang2024chemllm,ying2024internlm,touvron2023llama,achiam2023gpt}). In DocReLM, the LLM is utilized in two distinct ways. Initially, it functions as an automatic data annotator, generating high-quality training data to enhance retrieval models. This resulted in superior performance of our dense retriever and reranker models compared to other competitors, as illustrated in \ref{tab:reranker}. The performance of other models is not as good as ours since they are trained with general semantic similarity tasks. By training with data generated by LLM, our model is adapted to the document retrieval task. And the domain specialized knowledge is also distilled into the model in this process. 

Subsequently, LLM acts as a searching agent, refining paper selection based on textual information and reference relationships in the retrieved documents. This process is similar to the human reasoning process, where the user reads the references in the paper to find the most relevant ones. The LLM's ability to understand the context and relationships between references enables it to identify the most relevant papers, thereby enhancing the system's performance. We find this especially useful in the natural science domain, where papers are highly interconnected and understanding the references is crucial to identifying relevant papers. This is also a novel approach to implement the Large Language Model in the retrieval system.

The whole system of DocReLM is illustrated in Fig. \ref{fig:architecture}. DocReLM can be basically divided into three parts. The first two parts are a retriever and a reranker. These two models can filter and sort the document by the relevance to the user's query. The third model is a reference extractor that reads the retrieved documents and finds reference paper ID that can better answer the question. The extracted references are then added to the final document list. 

\begin{figure}[htp]
    \centering
    \begin{subfigure}{.5\textwidth}
        \centering
        \includegraphics[width=.9\linewidth]{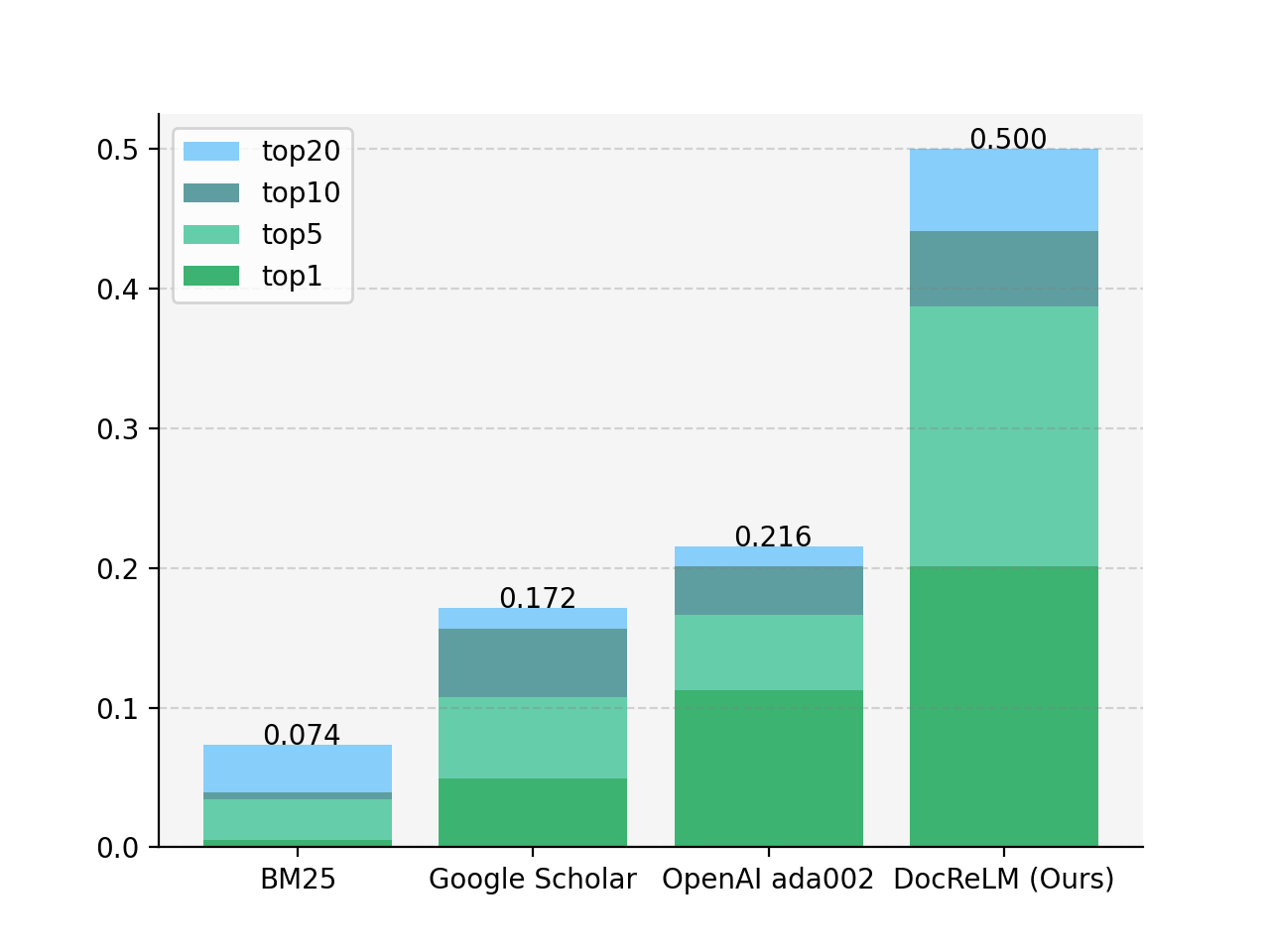}
        \caption{computer vision}
        \label{fig:acc-quantum}
    \end{subfigure}%
    \begin{subfigure}{.5\textwidth}
        \centering
        \includegraphics[width=.9\linewidth]{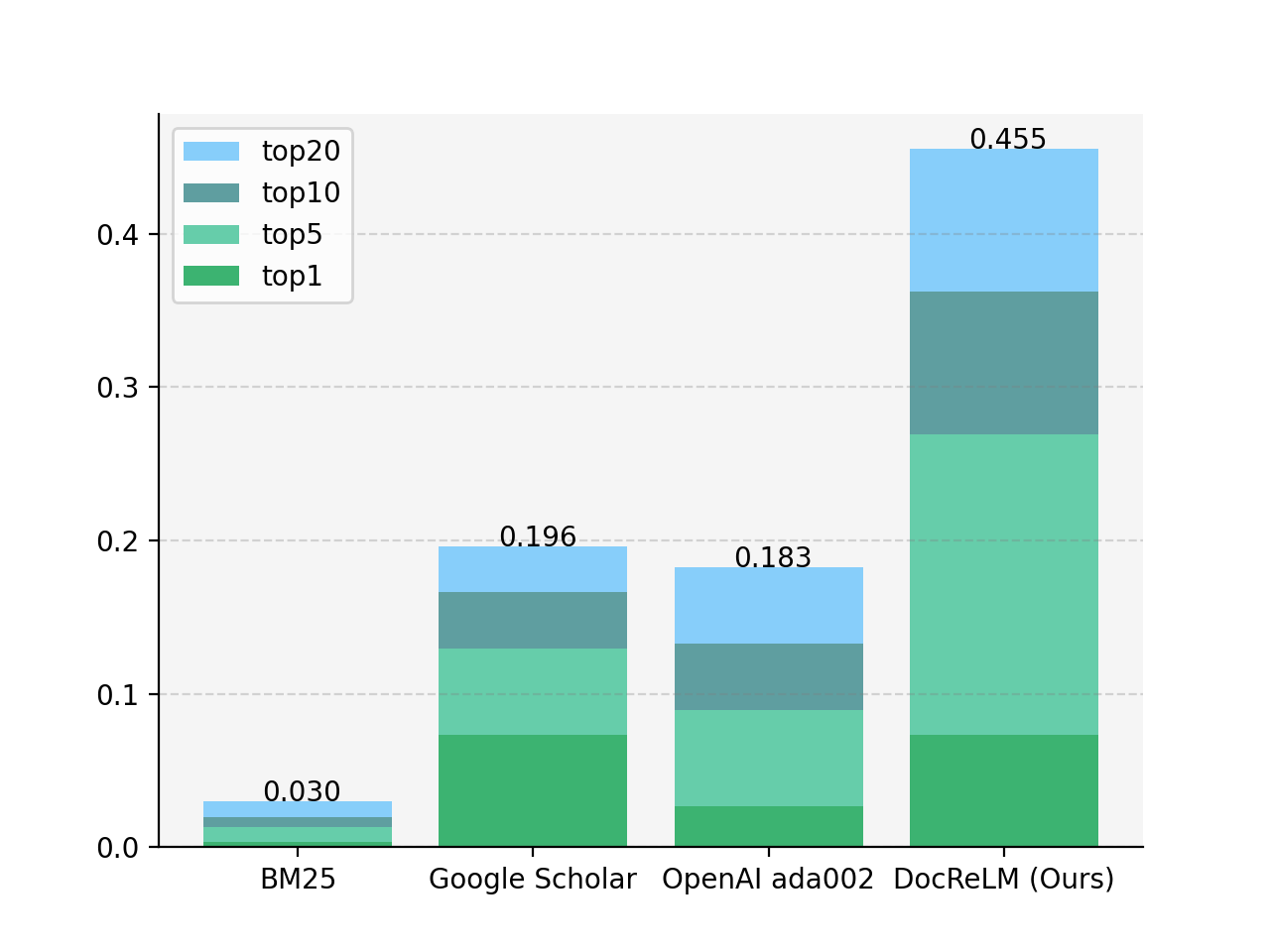}
        \caption{quantum physics}
        \label{fig:acc-cv}
    \end{subfigure}
    \caption{Accuracy of the final system}
    \label{fig:acc}
\end{figure}

To evaluate DocReLM, we established a benchmark for semantic document retrieval and conducted comparisons with several competitors. This benchmark is structured into two distinct tracks: quantum physics and computer vision. Each track encompasses a curated set of articles alongside a series of questions, with each question paired with a standard answer annotated by an academic researcher. In our experiments, DocReLM demonstrated a significant edge over its competitors, achieving an accuracy rate of 38.73\% in computer vision and 26.91\% in quantum physics on the top 5 results. The full result is depicted in \ref{fig:acc}. This highlights the system's potential to revolutionize document retrieval in academic research.

\section{System Architecture}\label{sec_architecture}
\begin{figure}
    \centering
    \includegraphics[width=1\linewidth,height=9cm]{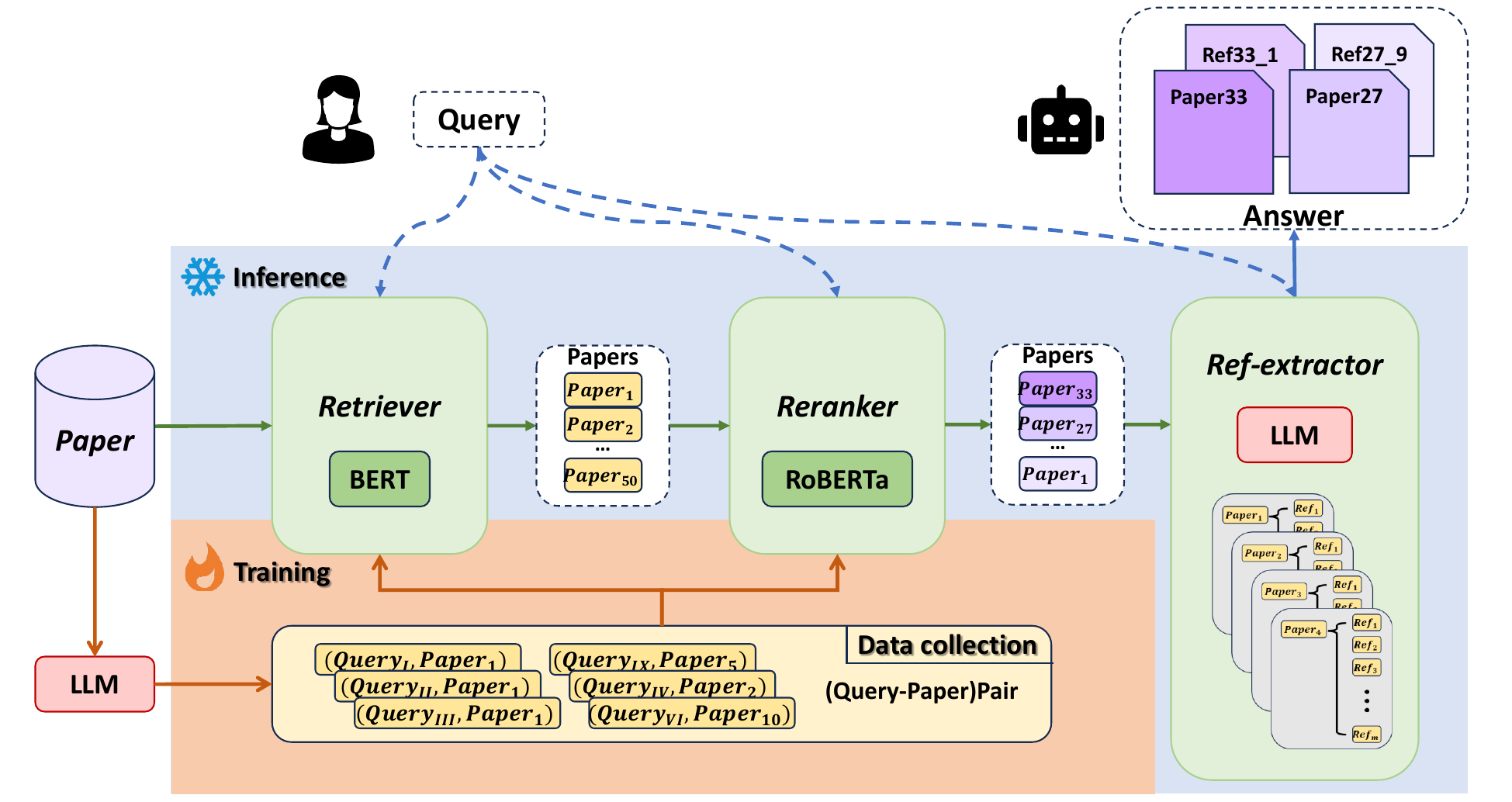}
    \caption{The training and inference of DocReLM. The LLMs are used in both training and inference.}
    \label{fig:architecture}
\end{figure}

The workflow of training and inferring DocReLM is depicted in Fig. \ref{fig:architecture}. The system comprises three primary components: retriever, reranker, and reference extractor. When presented with a user query, the retriever component utilizes embedding to fast retrieve candidate passages from the corpus. Subsequently, a reranker is employed to sort these passages more accurately. The final component examines the content of the top $k$ results provided by the reranker and generates appropriate references based on the contents of these documents. In the following subsections, we elaborate on the structural intricacies of each component in the model and how we utilize large language models to build these components. 

\subsection{Retriever}

The retriever is designed to efficiently extract a select set of document from a vast corpus. It requires rapid and effective retriever capabilities. Typically, two models are employed for this purpose: sparse retriever and dense retriever(\cite{10.1145/3486250}). The sparse retriever, such as the famous BM25 (\cite{bm25}), utilizes bag-of-words vector to match candidates. Conversely, dense embedding models explicitly encode sentences into dense vectors(\cite{karpukhin-etal-2020-dense}; \cite{gao-callan-2022-unsupervised}), employing cosine similarity to evaluate the likelihood of a match between query and passage embedding. In DocReLM, we adopt the neural dense retriever and utilize the open-source jina-embedding-v2-base(\cite{gunther2023jina}) as the foundational model, which we further train with our custom data. This retriever operates as a encoder model(\cite{devlin-etal-2019-bert}; \cite{reimers-gurevych-2019-sentence}). Encoder has been applied to various tasks in the scientific domain(\cite{ji2021dnabert}; \cite{zhou2023dnabert}; \cite{liang2023rethinking}; \cite{gong4493250single}). Leveraging contrastive learning, dense retrievers generate embedding that emphasize key passage details, enabling the linkage of semantically similar sentences even when they use dissimilar terminology. However, pretrained embedding models lack specificity for document retrieval tasks(\cite{w-etal-2023-query}), often presenting a domain knowledge gap with domain-specific papers(\cite{wang-etal-2022-gpl}). To address this, we incorporate a large language model to automatically annotate pseudo queries with academic paper corpora, thereby refining the embedding model to align more closely with our task. In constructing this retrieval system using our tailored embedding model, we process the document corpus, segment long documents into passages, and convert them to embedding. These vectors are stored in the vector database Faiss (\cite{johnson2019billion, douze2024faiss}). When a user inputs a query, DocReLM employs the same retrieval model to derive the query's semantic embedding and selects candidate passages based on the cosine distance between document and query embedding.

\subsection{Reranker}

Following the rapid retrieval process, the reranker is employed to enhance the precision of the results. While it is more accurate in general, the reranker operates at a slower pace compared to a retriever. This characteristic underpins its typical usage post-retrieval, where the retriever effectively narrows down the selection by eliminating mostly irrelevant passages. In this task, we use cross-encoder(\cite{nogueira-etal-2020-document}) which processes inputs consisting of a concatenated sequence of thequery string and candidate passage from the retrieval subsystem, demarcated by a [SEP] token. The reranker employs attention mechanism to analyze the input, consequently synthesizing a comprehensive feature for the entire input sequence. Subsequently, a linear layer translates this feature into a scalar value, signifying the likelihood that the given passage correctly answers the query. By integrating the query and passage in its input, the reranker fosters increased interaction between the two, thereby enhancing its expressive capabilities and outperforming the retriever in effectiveness. A limitation of the cross-encoder is its inability to preprocess the document corpus, necessitating fresh inference for each new query. Therefore it cannot be used as a standalone system and requires a retriever to provide a small set of relevant candidates. In DocReLM, we utilize the open-source model XLM-RoBERTa-large(\cite{conneau-etal-2020-unsupervised}) as the base model, which we further train with our custom data as the method of LCE(\cite{gao2021lce}). The top 200 passages, as determined by the retriever, are subsequently fed into the reranker to output a score for each concatenated pair. These candidate passages are subsequently reordered based on their respective scores.

\subsection{Reference Extraction}

\begin{figure}[htbp]
    \centering
    \includegraphics[width=\linewidth]{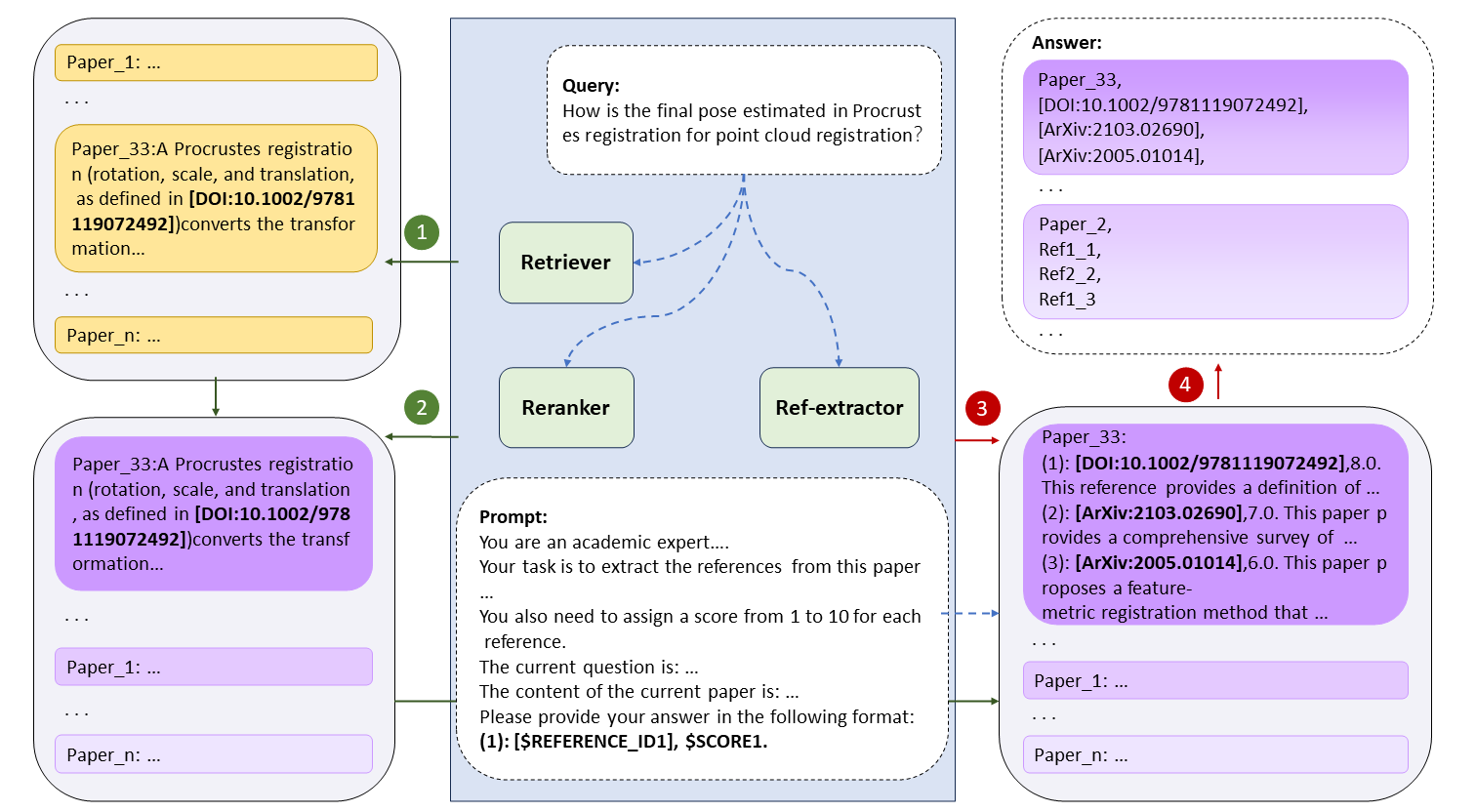}
    \caption{An Example for the reference extraction. A query is entered by a user. Retriever first return $n$ candidate passages from the entire corpus. Then the reranker sorts these passages. Finally, the reference extractor read the top 10 papers and extract three references for each of these paper, which is inserted after that original paper.}
    \label{fig:ref-extractor}
\end{figure}

After dense passage retrieval and cross-encoder reranking, the resulted papers are presented in the order of relevance with the query. We find that an additional procedure can improve the result significantly. That's the third part in DocReLM, reference extractor, which reads the passages content and extract the best references cited in the passage. This is aligned with practical searching. A paper chosen by retriever and reranker may not be the best paper to answer the query, but its reference lists has a large probability to contain that paper. 
Motivated by this objective, our approach involves instructing the large language model to extract the retrieved result and identify more suitable papers from the references. To achieve this, we manipulate the paper passage by inserting the identifier of the reference paper into the text.  When this modified passage is processed by a Large Language Model (LLM), it extracts these identifiers based on the contextual information. If the large language model determines that the passage itself constitutes the answer, it outputs the identifier of the referenced paper. Subsequently, our system conducts a search for the corresponding paper associated with the provided identifier. This progress is illustrated in Figure. \ref{fig:ref-extractor}. 

\section{Model and Training Details}\label{sec_model_and_train}
We use a two-stage training strategy to train the dense passage retriever and reranker in DocReLM. We first train the retriever with contrastive learning on the generated training data. Then we use the trained retriever to select hard negative samples for finetuning reranker with contrastive learning. In the following section, we discuss the data generation and training details.

\subsection{Generate Training Data with Large Language Model}
\label{section:data}

To train the retriever and reranker using contrastive learning, a collection of user queries and document passages is required. While datasets like MS MARCO are available with human-annotated labels (\cite{DBLP:conf/nips/NguyenRSGTMD16}), the expense of human annotation makes it challenging to acquire appropriate domain-specific retrieval datasets (\cite{wang-etal-2022-gpl}). In this work, we use a Large Language Model (LLM) to generate pseudo-queries from documents. We employ papers from unarXive (\cite{10.1007/s11192-020-03382-z}) to create the dataset, processing their original data to remove semantically irrelevant text and incorporating mathematical LaTeX and reference identifiers. These identifiers are formatted as \textit{Ref.X of IDF$_p$}, where \textit{IDF$_p$} is the identifier of the parent paper. We divide the entire paper into passages and use vicuna-7b-v1.5-16k (\cite{NEURIPS2023_91f18a12}) to generate a query for each sentence by providing the passage, title, and abstract. Inspired by the chain of thought approach (\cite{NEURIPS2022_9d560961}), we initially employ a zero-shot method (\cite{NEURIPS2022_8bb0d291}) to guide the LLM in creating an outline based on the title, abstract, and content, before generating the queries.

We create two subsets for the experiment. One is the quantum physics category. It has $56927$ papers and we collect $2.8M$ training data. The other is the computer vision category. It has $37390$ papers and we collect $1.1M$ training data. These data are both used to train the dense passage retriever and reranker. 

\subsection{Contrastive learning}
We follow the common strategy to training the retriever and reranker with contrastive learning. With the data generated in Sec \ref{section:data}, we have multiple pairs of query and passage, where the query $Q$ and passage $P$ in the same pair pairs are labeled positive, and that in different pairs are labeled negative. Contrastive learning can be seen as a binary classifier trained these pairs. For each positive pair $(Q_i, P_i)$, we sample $N$ negative pairs $(Q_i, P_{i,n})$ and compute the loss using Eequation \ref{simcse}. 
\begin{equation}
    \ell_i=-\log \frac{\mathrm{e}^{s(Q_i,P_i) / \tau}}{\mathrm{e}^{s(Q_i,P_i) / \tau} + \sum_{n=1}^N \mathrm{e}^{s(Q_i,P_{i,n}) / \tau}},
    \label{simcse}
\end{equation}
where $s(Q,P)$ denotes the score function for $(Q, P)$. In retriever, $s(Q, P)$ is given by the cosine similarity of the embeddings of $Q$ and $P$. In reranker, $s(Q, P)$ is directly given by the cross-encoder. 
By minimizing Equation \ref{simcse}, the positive pair scores are optimized to be significantly larger than the negative pair scores. 




\subsection{Negative Pairs for Retriever and Reranker}
The negative pairs in contrastive training retriever and reranker are generated differently. For retriever, these negative data are randomly sampled. In each training batch, $N+1$ positive pairs are sampled, each of which has $N$ negative pairs in the batch. \cite{neelakantan2022text} find larger batch size can improve the retriever's performance by increasing the chance of introducing hard negative. We use gradient cache technique(\cite{gao-etal-2021-scaling}) to increase batch size for better performance in contrastive learning. 

For training reranker, we select negative data with our trained retriever as the method of \cite{gao2021lce} called LCE. For each query $Q_i$, the retriever return $N$ passages that have the highest scores. These passages are highly related to the query. This means we utilized the most confusing wrong passages apart from the correct one. Therefore, these wrong passages are used as high quality hard negative samples for training the reranker. 


\section{Evaluation and results}
In assessing our document retrieval system, we set up a benchmark with queries taken from actual research scenarios. We invite researchers to create queries in their research domains and provide papers as annotated labels. To ensure the reliability of these annotations, the researchers are required to provide claims from surveys. Subsequently, we create a subset of unarXive as the corpus for retrieval. The final benchmark comprises two tracks: the quantum physics track, which includes 301 queries and 56927 papers in the corpus, and the computer vision track, which consists of 204 queries and 37390 papers in the corpus. We compare the performance of our retrieval system with that of Google Scholar. Furthermore, we conduct an ablation study to evaluate the individual effectiveness of the three components in our system. We highlight their advantages over similar models and demonstrate the superiority of our model compared to pre-trained models.

\begin{table}[h]
 \centering
 \begin{tabular}{llllll}
  \toprule
  Domain & Method & top-1 & top-5 & top-10 & top-20 \\
  \midrule
  computer vision & BM25  & 0.49 & 3.43 & 3.92 & 7.35 \\
  computer vision & text-embedding-ada-002  & 11.27 & 16.67 & 20.09 & 21.57 \\
  computer vision & jina-base-v2 & 8.33 & 14.71 & 18.14 & 22.55 \\
  computer vision & DocReLM-retriever & \textbf{17.16} & \textbf{30.39} & \textbf{39.22} &  \textbf{43.63} \\
  \midrule
  quantum physics & BM25  & 0.33 & 1.33 & 1.99 & 2.99 \\
  quantum physics & text-embedding-ada-002  & 2.66 & 8.97 & 13.29 & 18.27 \\
  quantum physics & jina-base-v2 & \textbf{3.99} & 9.97 & 14.62 & 18.27 \\
  quantum physics & DocReLM-retriever & 3.32 & \textbf{11.96} & \textbf{15.95} &  \textbf{21.93} \\
  \bottomrule
 \end{tabular}
 \caption{Results of the retriever.}
 \label{tab:embedding}
\end{table}

We compare three baseline retrieval models for this task: BM25, ada-002 from OpenAI and jina-base-v2. The accuracy is shown in Table \ref{tab:embedding}. The sparse retriever, BM25, shows poor performance compared to other models. Word frequency is insufficient to capture semantic relation in this task.  Our DocReLM-retriever outperforms all of these models in both domains.  In the computer vision domain, DocReLM-retriever significantly outperforms the other models. It achieves a top-1 accuracy that is 5.89\% higher than the best embedding model, and top-5 and top-10 accuracy that are 13.72\% and 19.13\% higher, respectively. It is worth noting that all three models perform well in the computer vision domain, with high average accuracy. In the quantum physics domain, the performance of the embedding models is similar, with no significant difference in the top-1 results. However, when considering the top-5 results, our model improves the accuracy from 9.97\% to 11.96\% to the best of other models. 

\begin{table}[h]
 \centering
 \begin{tabular}{llllll}
  \toprule
  Domain & Method & top-1 & top-5 & top-10 & top-20 \\
  \midrule
  computer vision & without rerank  & 17.16 & 30.39 & 39.22 & 43.63\\
  computer vision & bge-reranker-large  & 19.61 & 35.78 & 41.67 & 44.12\\
  computer vision & Cohere rerank-english-v2.0 & 16.66 & 34.31 & 40.20 & 43.63  \\
  computer vision & DocReLM-reranker & \textbf{20.10} & \textbf{37.75} & \textbf{44.61} & \textbf{47.06}	\\
  \midrule
  quantum physics & without rerank  & 3.32 & 11.96 & 15.95 & 21.93 \\
  quantum physics & bge-reranker-large & 5.32 & 14.29 & 20.27 & 25.25\\
  quantum physics & Cohere rerank-english-v2.0 & \textbf{7.97} & \textbf{17.28} & \textbf{21.93} & 25.91 \\
  quantum physics & DocReLM-reranker & 7.31 & \textbf{17.28} & 19.93 & \textbf{26.58} \\
  \bottomrule
 \end{tabular}
 \caption{Results of the reranker based on the embedding of jina-base-v2-tuned.}
 \label{tab:reranker}
\end{table}

To test our reranker's performance, we choose three different rerankers, bge-reranker-large(\cite{xiao2023c}), Cohere, and DocReLM-reranker to compare. The rerankers take the top 200 passages selected by DocReLM-retriever as input data. The results are shown in Table \ref{tab:reranker}. In the computer vision domain, the DocReLM-reranker outperforms both the other models. It achieves a top-1 accuracy that is 0.49\% higher than the best reranker, and top-5 and top-10 accuracy that are 1.97\% and 2.94\% higher, respectively. In the quantum physics domain, the DocReLM-reranker and Cohere achieve competing performance. Our models perform better at top-20, while Cohere models excel at top-1 and top-10 scores. These results demonstrate that the DocReLM-reranker is more effective than other models in both domains. It highlights the importance of the reranker in improving the accuracy of the retrieval system.

\begin{table}[h]
 \centering
 \begin{tabular}{llllll}
  \toprule
  Domain & Method & top-1 & top-5 & top-10 & top-20 \\
  \midrule
  computer vision & without ref extraction & 20.10 & 37.75 & 44.61 & 47.06 \\
  computer vision & internLM & 20.10 & 38.73 & 44.12 & 50.00 \\
  \midrule
  quantum physics & without ref extraction & 7.31 & 17.28 & 19.93 & 26.58 \\
  quantum physics & internLM & 7.31 & 26.91 & 36.21 & 45.51 \\
  \bottomrule
 \end{tabular}
 \caption{Results of reference extraction.}
 \label{tab:ref-extraction}
\end{table}

Our last experiment evaluates the effectiveness of the reference extraction model. We compare the performance of the DocReLM-reranker with and without reference extraction. The results are shown in Table \ref{tab:ref-extraction}. In the computer vision domain, the reference extraction model internLM improves the top-5 and top-20 accuracy by 0.98\% and 2.94\%, respectively. These results demonstrate that the reference extraction model is effective in improving the accuracy of the retrieval system. In the quantum physics domain, the reference extraction model internLM improves the top-5 and top-10 accuracy by 9.63\% and 16.28\%, respectively.

\section{Discussion}
This paper presented DocReLM, an innovative approach to document retrieval using Large Language Models (LLM), demonstrating significant improvements over traditional retrieval methods. Our system integrates a neural dense retriever, a reranking mechanism, and a novel reference extraction component, all fine-tuned and enhanced through the capabilities of LLMs. In the subsequent subsections, we will discuss two aspects of DocReLM in greater detail.

\subsection{Distilling of LLM}
The neural dense retriever and cross-encoder reranker effectively capture semantic relations between queries and documents, facilitating more accurate retrieval of relevant document. We notice that compared to the base model we used, there is a significant improvement after training the model with our generated data. Since these data are generated with LLM, the comprehension ability of LLM is distilled into the relative small model, which can adapt a general model to a specific task. 

\subsection{combination of LLM and retrieval system}
This study also introduces a novel approach to integrating LLMs with retrieval systems. With the growing popularity of LLMs, there have been various efforts to merge them with search engines. These attempts have primarily followed two approaches: using LLMs to expand search queries or to summarize search results. However, DocReLM pioneers a new direction by enabling an LLM to understand search results and continue the search process, suggesting better candidates. This concept could be expanded further. The process could be iteratively repeated, allowing the LLM to traverse multiple nodes within a citation network until it identifies the most relevant results. Such a task requires the LLM to grasp the logical relationships between referenced papers and possess sufficient knowledge of the field to accurately extract references. By incorporating an LLM that can search multiple times and analyze results to refine its searches, a general retrieval system could also be significantly enhanced. This approach represents a promising direction for the future of retrieval systems.

\section{Conclusion}\label{sec_conclusion}
In conclusion, DocReLM appears to contribute positively to the field of academic document retrieval. This study, by integrating Large Language Models (LLMs) with traditional retrieval systems, seeks to address the challenges posed by the increasing volume and complexity of academic document and proposes a new direction in the application of LLMs for semantic-based search. The use of LLMs in DocReLM to generate high-quality pseudo data, enhance retriever and reranker models, and extract references from retrieved papers, represents a promising step forward in leveraging the rich information available in academic documents.

Our results suggest a notable improvement in retrieval accuracy compared to existing systems. This improvement may be attributed to the retrieval data generated by LLM for fine-tuning models.
Moreover, DocReLM introduces an innovative approach to employing LLMs in retrieval systems, demonstrating the potential of these models to navigate through multiple citation network nodes to locate relevant papers. This capability may signify a shift from traditional keyword-based searches to more nuanced, context-aware methods that reflect human reasoning processes.

As the landscape of academic research continues to evolve, the approach adopted by DocReLM could pave the way for developing more sophisticated, effective, and user-friendly document retrieval systems. The encouraging results of this study imply considerable potential for future enhancements, especially as LLMs further evolve and become increasingly capable of handling complex tasks in specialized domains.

\section{Acknowledgements}
We sincerely thank Dongzhan Zhou, Guoqiang Jin, and Xingyu Zeng for their invaluable discussions and constructive feedback. We also extend our gratitude to Lei Bai, Hao Du, Tianxiang Gao, Jingwen He, Tong He, Di Huang, Xiaoshui Huang, Meng Li, Yan Lu, Peixia Li, Wentao Qu, Shixiang Tang, Yuan Wang, Xiaopei Wu, Honghui Yang, Hongliang Yan, Peng Ye, Weicai Ye, Sha Zhang, and Peixin Zhuang for their diligent work in annotating the test dataset. 
This work is partially supported
 by the National Key R\&D Program of China
 (NO.2022ZD0160100), and in part by Shanghai
 Committee of Science and Technology (Grant No.
 21DZ1100100).

\newpage

\end{CJK}
\end{document}